\documentclass[3p,times,procedia]{elsarticle}
\usepackage{nupha_ecrc}

\volume{00}

\firstpage{1}

\journalname{Nuclear Physics A}

\runauth{}

\jid{nupha}

\jnltitlelogo{Nuclear Physics A}

\usepackage{amssymb}

\usepackage[figuresright]{rotating}
\usepackage{xspace}

\newcommand{\pt}{$p_{\rm T}$\xspace}

\begin{document}

\begin{frontmatter}



\dochead{}

\title{PHENIX results on centrality dependence of yields and correlations
in $d$+Au collisions at $\sqrt{s_{_{NN}}}$=200\,GeV}


\author{Takao Sakaguchi, for the PHENIX Collaboration\footnote{For the full PHENIX Collaboration author list and acknowledgments, see Appendix ``Collaboration'' of this volume}}

\address{Brookhaven National Laboratory, Physics Department, Upton, NY 11973, USA.}

\begin{abstract}
PHENIX has measured the transverse momentum (\pt) spectra and two particle
angular correlations for high \pt particles in $d$+Au collisions
at $\sqrt{s_{_{NN}}}$=200\,GeV using the RHIC Year-2008 run data. The azimuthal
angle correlations for two particles with a large rapidity gap
exhibit a ridge like structure. Using the $\pi^0$ reconstructed in the EMCal,
we have successfully extended the \pt reach of the correlation up to 8\,GeV/$c$.
We find that the azimuthal anisotropy of hadrons found at low \pt persists
up to 6\,GeV/$c$ with a significant centrality and \pt dependence, similar to
what was observed in A+A collisions.
\end{abstract}


\end{frontmatter}


\section{Introduction}
The small collision systems such as $p/d$+A collisions have been
considered a good laboratory to quantify cold nuclear matter effects,
a necessary baseline for understanding the effects of the hot and
dense medium produced in A+A collisions. After the the ridge-like
structure in the long-range rapidity correlation in $p$+Pb
collisions at $\sqrt{s_{_{NN}}}$=5.02\,TeV at the LHC was
reported~\cite{Aad:2012gla, Aad:2014lta},
however, the systems can no longer be considered as a simple cold
nuclear matter. The study at the LHC was followed by the PHENIX
experiment at RHIC, and a finite $v_2$ of hadrons in
0-5\,\% central $d$+Au collisions using both the two-particle angular
correlation method and the event-plane method were
shown~\cite{Adare:2013piz, Adare:2014keg}.
These observations led the community to explore any phenomena found
in A+A collisions, in $p/d$+A
collisions. Recently, interest has focused on the high \pt region
in the small systems. If the collective behavior in low \pt is
hydrodynamic, the phenomena should cease at higher
\pt~\cite{Bozek:2013uha, Bzdak:2013zma}, while
a CGC-motivated model would produce a correlation even at
high \pt~\cite{Dusling:2013qoz}. The hydrodynamical scenario
would also suggest that the energy loss of hard scattered partons
may occur in the small systems, and result in small
but sizable anisotropy of the particle emission at high \pt.
The recent PHENIX measurement of the reconstructed jets in 
minimum bias $d$+Au collisions at $\sqrt{s_{_{NN}}}$=200\,GeV shows
little or no modification of their production rates compared to those
expected from p+p collisions within the experimental
uncertainty, while a strong centrality dependence
in the rates has been observed~\cite{Adare:2015gla}.
Extending the measurement of $v_n$ (or $c_n$) to high \pt may
contribute to understanding the interplay of high \pt particles
and the possible medium created in the $d$+Au collisions.
We show the latest results on the azimuthal angle correlation of
pairs of hadrons with a large rapidity gap
in $d$+Au and $p+p$ collisions.

\section{Analysis}
\begin{figure}[htbp]
\begin{minipage}{84mm}
\centering
\includegraphics[width=8.2cm]{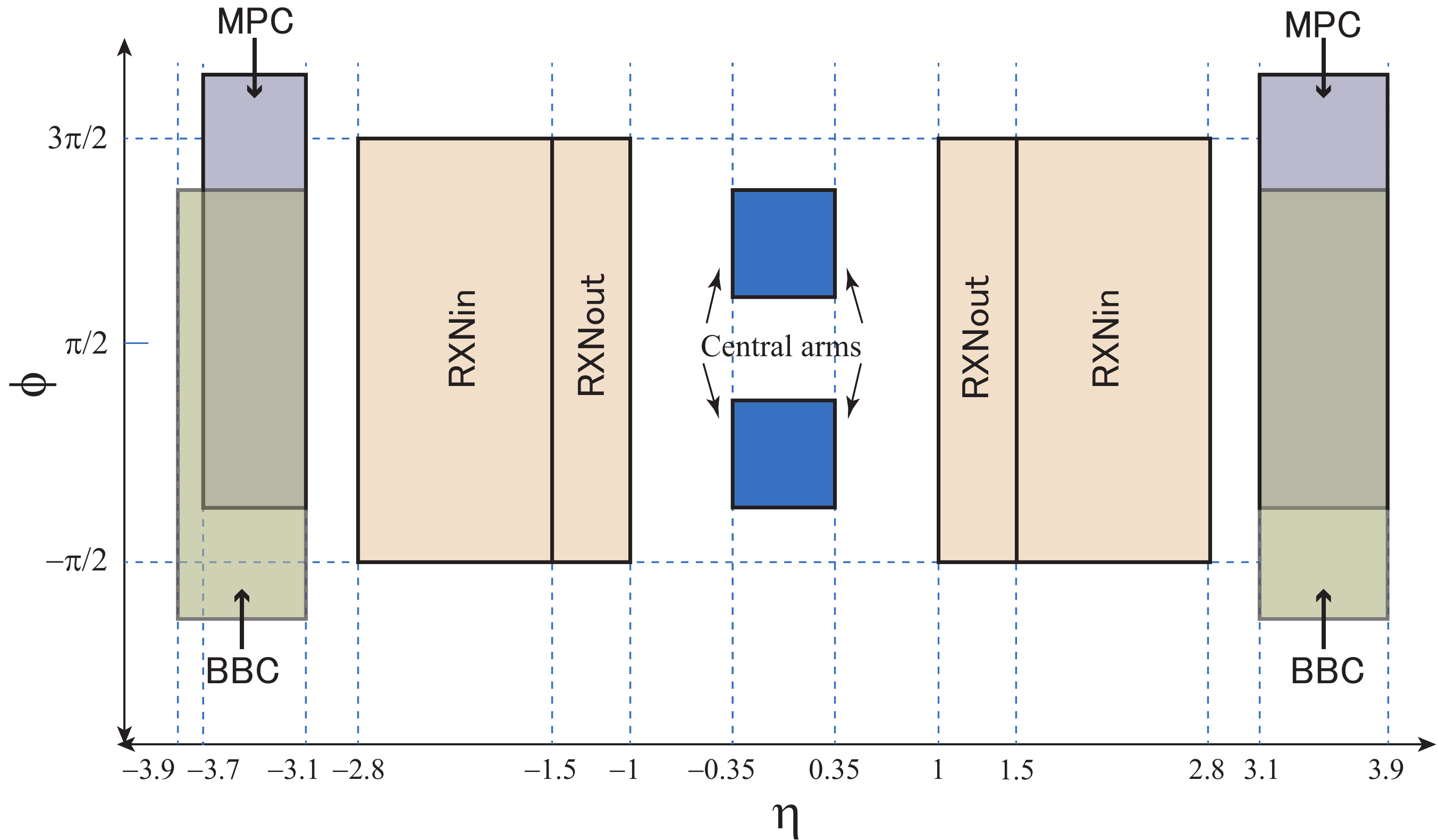}
\end{minipage}
\hspace{2mm}
\begin{minipage}{64mm}
\centering
\includegraphics[width=6.3cm]{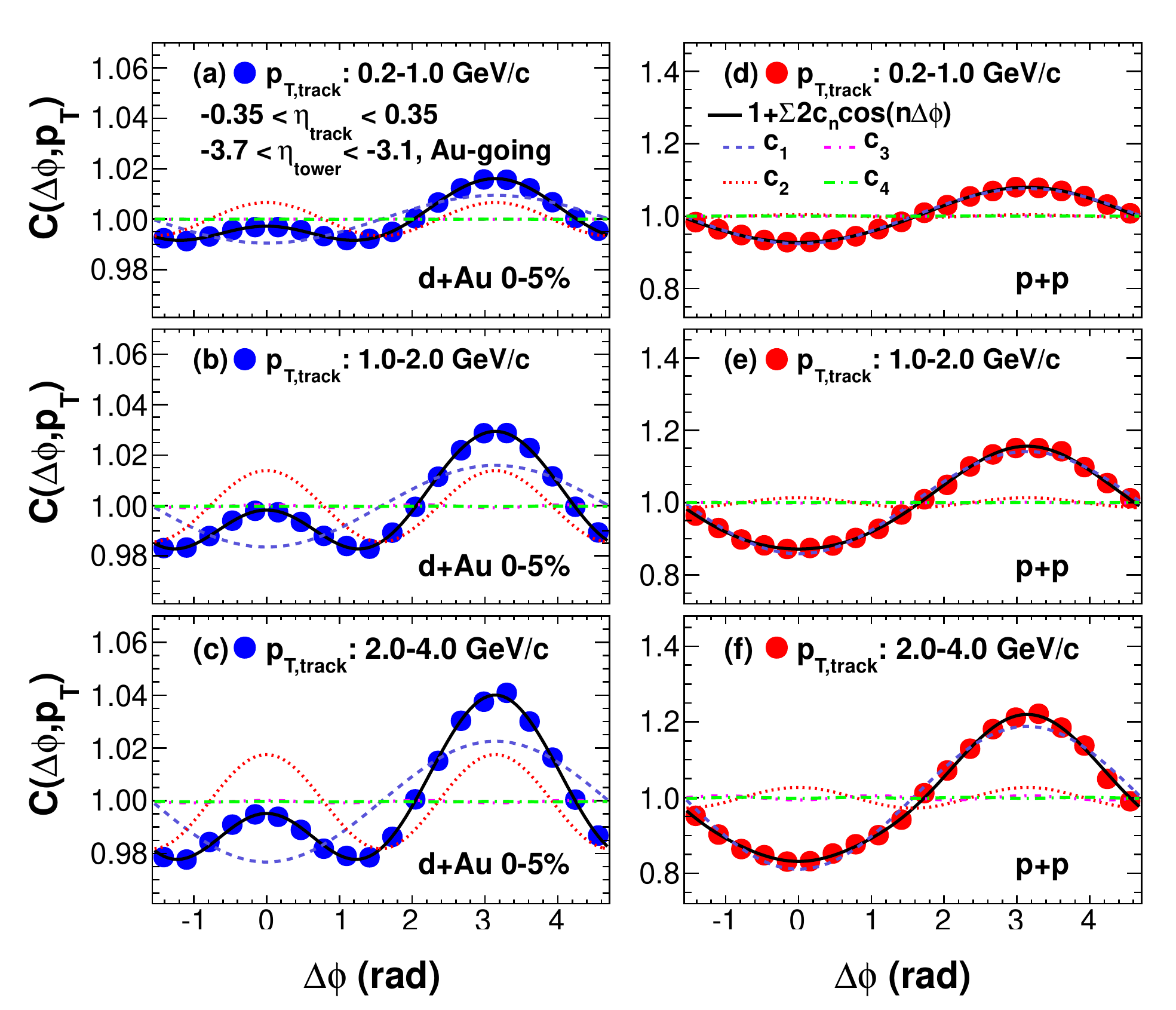}
\end{minipage}
\caption{ (a, left) Configuration of PHENIX detector in $\phi-y$
coordinate.  (b, right) Two particle angular correlation functions
using charged hadrons measured in CNT, and tower energies measured
in MPC south detector in 0-5\,\% $d$+Au collisions.}
\label{fig1}       
\end{figure}

PHENIX recorded an integrated luminosity of 80\,nb$^{-1}$ in
$d$+Au collisions and that of 5.2\,pb$^{-1}$ in $p$+$p$ collisions
in RHIC Year-8 run. The events were triggered by Beam-Beam
counter (BBC) located at 3.1$<|\eta|<$3.9 covering the full
azimuth. The detail description of the PHENIX detector system
is found in a literature~\cite{Adcox:2003zm}.
The long-range two particle correlation functions were constructed
by pairing the charged hadrons in the mid-rapidity region
($|\eta|<$0.35, defined as CNT) with the energy deposited in the
towers of Muon Piston Calorimeter (MPC) sitting either south
(-3.7$<\eta<$-3.1) or north (3.1$<\eta<$3.9). Figure~\ref{fig1}(a)
show the position of each PHENIX detector component in the
$\phi-y$ coordinate.
Note that the MPC south (MPCS) is sitting in Au-going direction,
which has more multiplicity, in case of $d$+Au collisions, while
the MPC north (MPCN) is sitting in $d$-going direction.
For reaching higher \pt, we substituted the charged hadrons with
$\pi^0$ measured by the electromagnetic calorimeter (EMCal)
through the $\pi^0\rightarrow \gamma\gamma$ channel.
We defined the $\gamma\gamma$ invariant mass region of
0.12-0.16\,GeV/$c^2$ as the $\pi^0$ region. There is a combinatoric
background underneath the $\pi^0$ peak, and we accounted for this
contribution in systematic uncertainty estimate.
We applied different energy threshold for MPC tower energy
between charged hadrons and $\pi^0$; we used 3\,GeV when
associating with charged hadrons and 0.3\,GeV for $\pi^0$.

After constructing correlation functions, we fitted them with
Fourier series for quantifying their shape:
\[\frac{dN}{d\Delta\phi} = N_0{1+2c_1 cos(\Delta\phi)+2c_2 cos(2\Delta\phi)+2c_3 cos(3\Delta\phi)+2c_4 cos(4\Delta\phi)}\]
where $c_n$ can be written as $c_n =v_n(MPC)\times v_n(CNT)$.
In the following section, we show the correlation functions as well as
the $c_n$'s obtained from the fit.

\section{Results and discussion}
\begin{figure}[htbp]
\begin{minipage}{8.5cm}
\centering
\includegraphics[width=8.4cm]{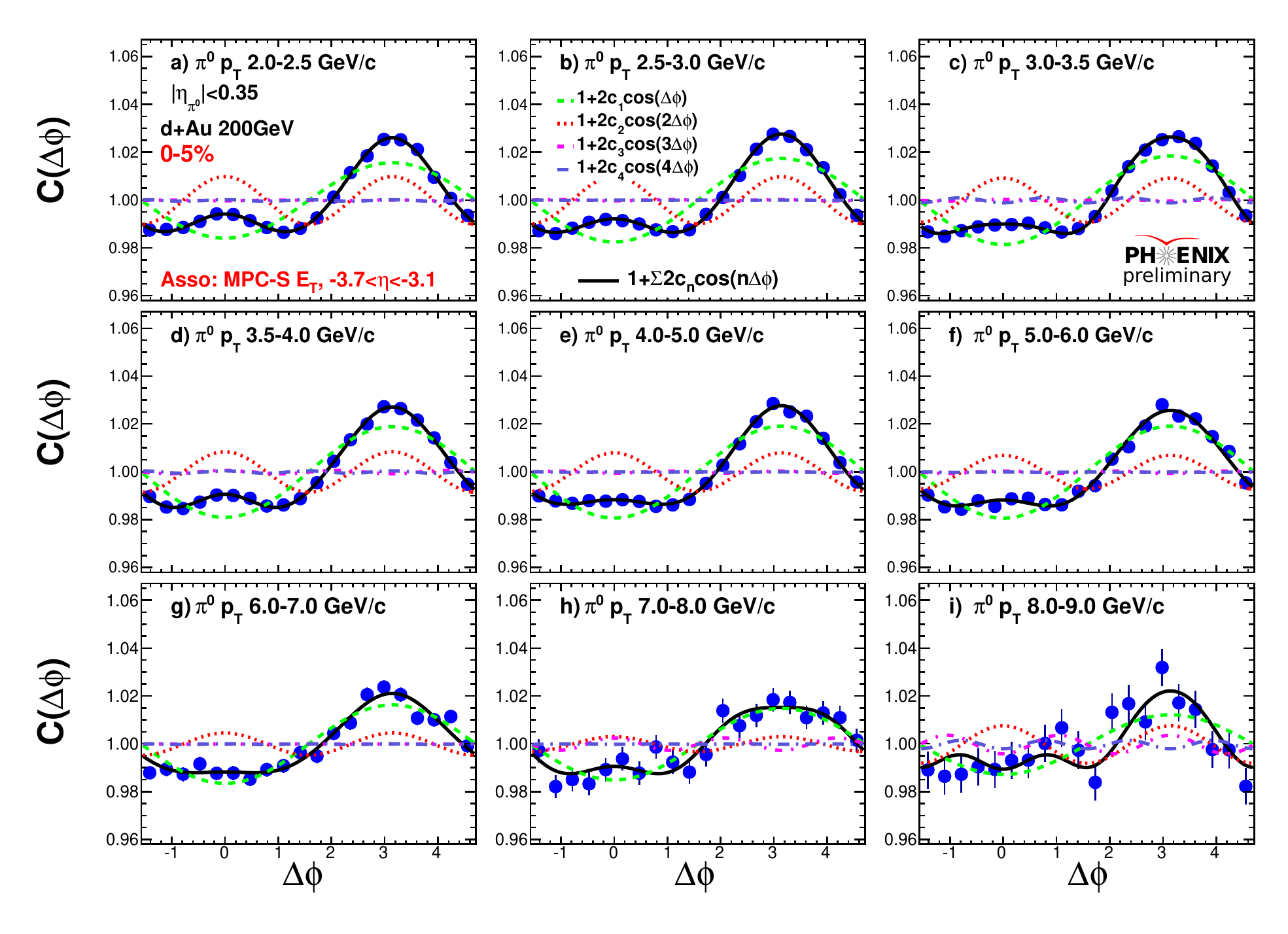}
\end{minipage}
\hspace{2mm}
\begin{minipage}{6.3cm}
\centering
\includegraphics[width=6.1cm]{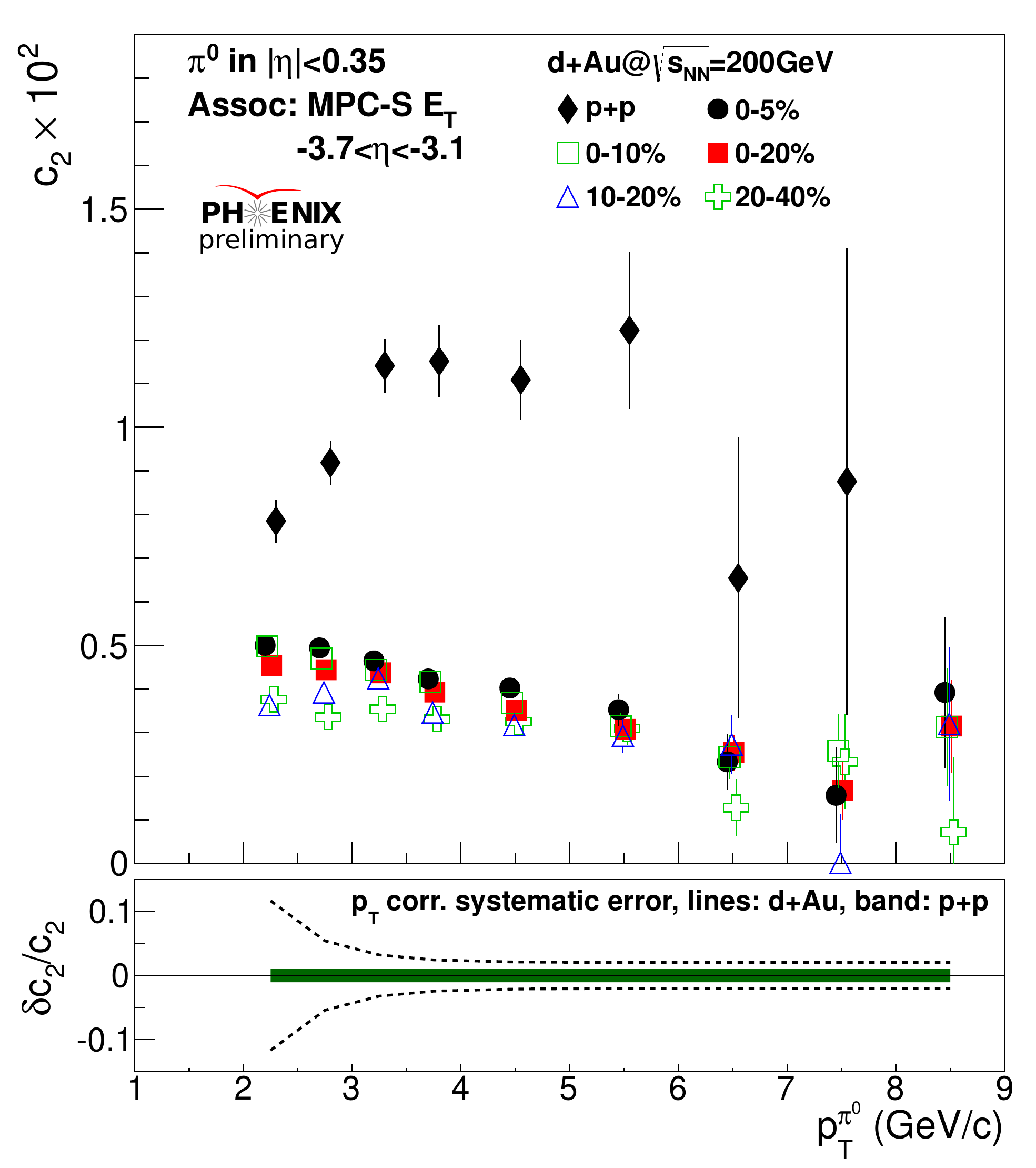}
\end{minipage}
\caption{(a, left) Two particle angular correlation functions using $\pi^0$
reconstructed in CNT, and tower energies measured in MPC south
detector in 0-5\,\% $d$+Au collisions. (b, right) $c_2$ values obtained from Fourier series fit to the
$\pi^0$-MPCs correlation functions.}
\label{fig2}       
\end{figure}

In Figure~\ref{fig1}(b), we show the correlation functions for
charged hadrons in CNT associated with MPCS for 0-5\,\% $d$+Au
collisions, together with the ones in $p+p$ collisions as reference.
Comparing to the $p+p$ collision case, the correlation functions
for $d$+Au collisions have peaks at near side ($\Delta\phi\sim$0),
which is well described by a second order Fourier term. The peaks
become much
prominent as going to higher \pt. It was found that the similar
correlation is seen for CNT-MPCN correlation for the same centrality,
but the magnitude is much smaller. Solid lines show the total fit
to the Fourier series as described in the previous section, while
the dotted lines show each term of the series.
Using $\pi^0$ in CNT, we have successfully extended our measurement
in \pt compared to that from charged hadrons as shown in
 Figure~\ref{fig2}(a).
The peaks at near side are clearly seen up to \pt$\sim$6\,GeV/$c$,
while the magnitudes decrease with \pt.
The definition of the lines are same as Figure~\ref{fig1}(b).
In order to quantify the line shape of the correlation functions,
we plotted the $c_2$ from the Fourier series fit for various
centralities in $d$+Au collisions as a function of \pt of $\pi^0$,
together with that in $p+p$ collisions. The result is shown in
Figure~\ref{fig2}(b).
The bottom panel show the relative systematic uncertainty for the
data points shown in the top panel. The main component of the
uncertainty is from the uncertainty of the correlation between
combinatoric background under $\pi^0$ and MPCS.
As was seen in the correlation functions, the $c_2$'s are decreasing
as a function of \pt. Due to the large magnitude of $c_2$ in $p+p$
collisions,
the $c_2$'s from $d$+Au collisions look similar. However, there
is a clear centrality dependence of the $c_2$; higher $c_2$ values
are seen as going to more central collisions.

\begin{figure}[htbp]
\begin{minipage}{73mm}
\centering
\includegraphics[width=6.1cm]{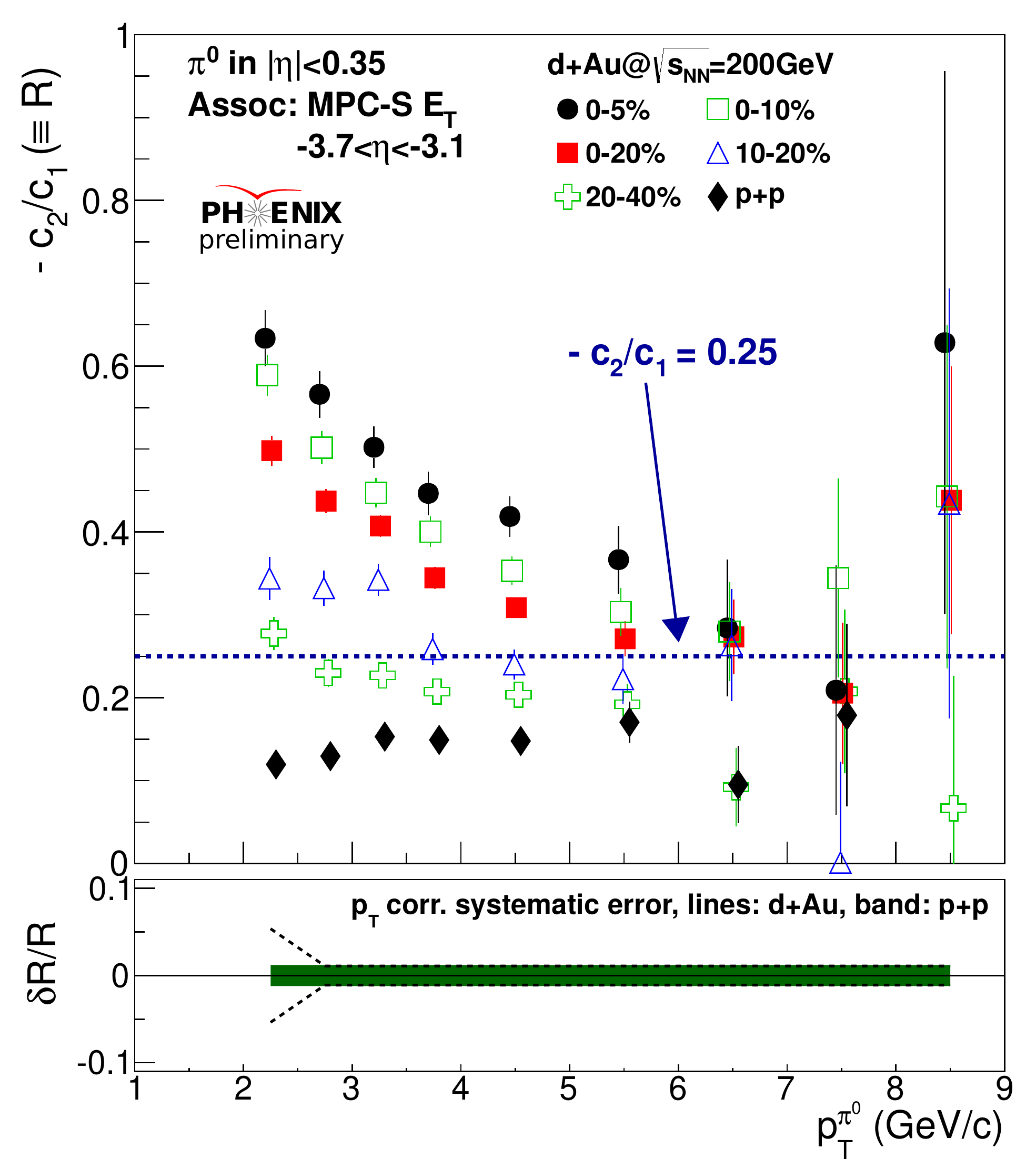}
\end{minipage}
\hspace{2mm}
\begin{minipage}{73mm}
\centering
\includegraphics[width=6.1cm]{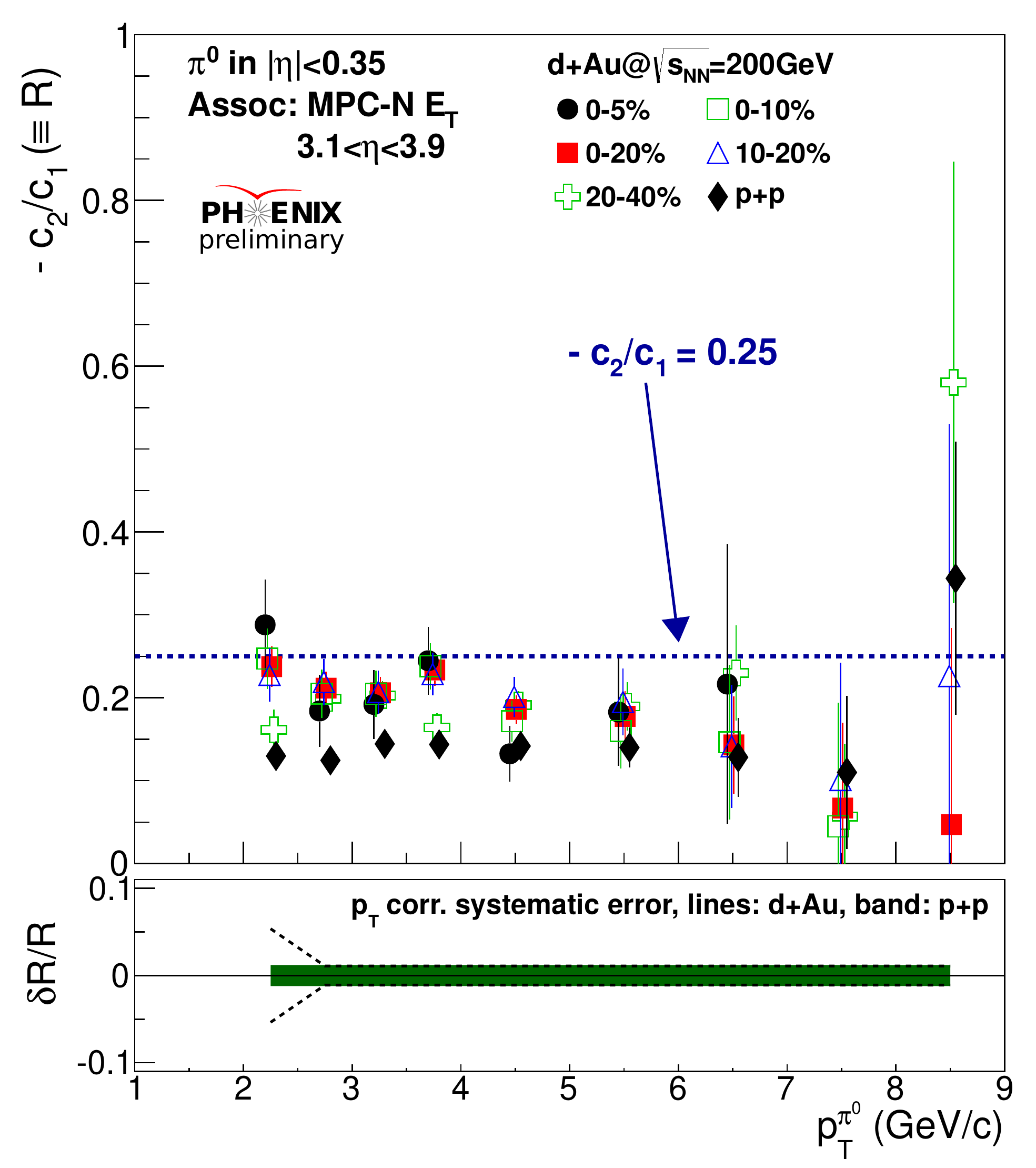}
\end{minipage}
\caption{ $-c_2/c_1$ values obtained from Fourier series fit to the
$\pi^0$-MPCS (a, left) and $\pi^0$-MPCN (b, right) two particle
angular correlation functions.}
\label{fig3}       
\end{figure}

The $c_1$ is the dipole component and primarily resulted in from
back-to-back jet contribution or energy conservation of the system.
Therefore, we assume that $c_1$ can be a proxy of primordial
multiplicity-related effect and ``normalize'' $c_2$ so that we can
compare apple-to-apple level among various centralities in $d$+Au
and $p+p$ collisions. With this idea in our mind, we computed
$-c_2/c_1$ as shown in Figure~\ref{fig3}(a).
The bottom panel is showing the systematic uncertainty as similarly
defined as in Figure~\ref{fig2}. Note that both statistical and systematic
uncertainties are smaller in $-c_2/c_1$ compared to $-c_2$, because some
of the uncertainties in $c_2$ and $c_1$ are correlated and can be
canceled out in the ratio. This way, we clearly see the centrality
dependent change of Fourier coefficients, which approach to
the $-c_2/c_1$ in $p+p$ collisions, as going to more peripheral collisions.
Assuming $c_3=c_4=0$ and requiring
$(\partial^2/\partial\Delta\phi^2) (dN/d\Delta\phi)=0$ result in 
$-c_2/c_1=0.25$ which is also shown on Figure~\ref{fig3}(a) as a
dotted line.
If the points are above this line, the distribution shows ``hill'',
while the distribution shows ``valley'' if the points are below.
We found that $c_3=c_4=0$ is a very good assumption in $d$+Au
and $p+p$ collision case. As we see in the correlation functions,
the hill persists up to $\sim$6\,GeV/$c$ for 0-5\,\% $d$+Au collisions.
Finally, we show the CNT-MPCN correlation for a comparison,
again using $\pi^0$ in CNT in Figure~\ref{fig3}(b). Although the
finite $-c_2/c_1$ and their slight centrality dependence are seen, 
the trend is very different compared to the one from CNT-MPCS.
This shows the ridge like structure is asymmetric in rapidity.
This result is consistent with previous PHENIX
results as well as the recent STAR publication~\cite{Adamczyk:2015xjc}.

These results show that the \pt and centrality dependence of the
ridge like structure in $d$+Au collision resembles what was
observed in A+A collisions, indicating the possibility that both
collective behavior and differential energy loss play a role also
in $d$+Au collisions. The result may contribute to understanding
the centrality dependence of the reconstructed jets~\cite{Adare:2015gla}.

\section{Summary}
In these proceedings, the latest results on two particle angular
correlations for high \pt particles in $d$+Au collisions at
$\sqrt{s_{_{NN}}}$=200\,GeV are
shown. We found that the azimuthal angle correlations for two
particles with a large rapidity gap exhibits a ridge like structure
mainly coming from the collective nature of the system. Using the
$\pi^0$ reconstructed in the EMCal, we have successfully extended
the \pt reach of the correlation up to 8\,GeV/$c$.  We found that
the azimuthal anisotropy of hadrons found in low \pt persists up
to 6\,GeV/$c$ with a significant centrality and \pt dependence,
similar to what was observed in A+A collisions.





\bibliographystyle{elsarticle-num}
\bibliography{Sakaguchi_QM15Proceedings_v3}







\end{document}